\documentstyle[12pt,axodraw]{article}
\def\be{\begin{equation}}
\def\ee{\end{equation}}
\def\bea{\begin{eqnarray}}
\def\eea{\end{eqnarray}}
\def\barr{\begin{array}}
\def\earr{\end{array}}
\newcommand{\sm}{standard model}

\newcommand{\xs}{cross section}
\newcommand{\lc}{linear collider}
\newcommand{\lk}{leptoquark}
\newcommand{\tenrm}{\mbox{}}

\begin{document}

\title{\Large\bf\boldmath 
LEPTOQUARK PRODUCTION IN $e^-e^-$ SCATTERING
}
\author{{\large FRANK CUYPERS}\\\\
{\tt cuypers@mppmu.mpg.de}\\
{\em Max-Planck-Institut f\"ur Physik, Werner-Heisenberg-Institut,}\\
{\em D-80805 M\"unchen, Germany}
}
\date{}
\maketitle
\pagestyle{empty}
\thispagestyle{empty}
\begin{abstract}
\noindent\footnotesize
We consider the production of scalar and vector leptoquarks 
in $e^-e^-$ collisions.
Although this reaction is not competitive
with {\em e.g.} $e^-\gamma$ scattering
for discovering these heavy states,
it is the main process susceptible of differentiating
two important classes of leptoquarks.
\end{abstract}

\bigskip\bigskip
%\bigskip
%\bigskip\bigskip\bigskip
%\vfill

\noindent
Leptoquarks appear in a wealth of extensions of the \sm,
ranging from grand unified to composite models.
A general classification of these states
which respects $SU(3)_c \otimes SU(2)_L \otimes U(1)_Y$ invariance
was performed in Ref.~\cite{brw}.
It assumes that:
\begin{itemize}
\item   
By definition,
  they couple to leptons and quarks.
  Therefore,
  they must be either singlets, doublets or triplets
  of the weak gauge group $SU(2)_L$.
\item
  For simplicity,
  their couplings to leptons and quarks
  do not involve derivatives.
  This ensures only lowest dimension operators are involved.
\item
  In order not to induce rapid proton decay 
  or other nuisances,
  they must conserve lepton $(L)$ and baryon $(B)$ number separately.
\end{itemize}
The most general effective lagrangian
which respects these conditions
reads~\cite{brw}
%can be separated into two parts,
%each involving 
%either \lk s which carry no fermion number $F=3B+L=0$,
%or \lk s with fermion number $F=2$ \cite{brw}:
\begin{eqnarray}
\label{lag}
L%{\cal L}
& = & 
\bigl( 
  h_{2L} \bar{u}_R \ell_L
+ h_{2R} \bar{q}_L i \sigma_2 e_R
\bigr) R_2
+ \tilde{h}_{2L} \bar{d}_R \ell_L \tilde{R}_2
+ h_{3L} \bar{q}_L \mbox{\boldmath$\sigma$} \gamma^\mu \ell_L \mbox{\boldmath$U$}_{3 \mu}
\\
& + &
\bigl( 
  h_{1L} \bar{q}_L \gamma^\mu \ell_L
+ h_{1R} \bar{d}_R \gamma^\mu e_R 
\bigr) U_{1 \mu}
+ \tilde{h}_{1R} \bar{u}_R \gamma^\mu e_R \tilde{U}_{1 \mu}
\nonumber\\
%& + &
%h_{3L} \bar{q}_L \mbox{\boldmath$\sigma$} \gamma^\mu \ell_L \mbox{\boldmath$U$}_{3 \mu}
%\\
& + & 
\bigl( 
  g_{1L} \bar{q}^c_L i\sigma_2 \ell_L
+ g_{1R} \bar{u}^c_R e_R
\bigr) S_1
+ \tilde{g}_{1R} \bar{d}^c_R e_R \tilde{S}_1
+ g_{3L} \bar{q}^c_L i\sigma_2\mbox{\boldmath$\sigma$} \ell_L \mbox{\boldmath$S$}_3
\nonumber\\
%& + &
%g_{3L} \bar{q}^c_L i\sigma_2\mbox{\boldmath$\sigma$} \ell_L \mbox{\boldmath$S$}_3
%\\
& + & 
\bigl( 
  g_{2L} \bar{d}^c_R \gamma^\mu \ell_L
+ g_{2R} \bar{q}^c_L \gamma^\mu e_R
\bigr) V_{2 \mu}
+ \tilde{g}_{2L} \bar{u}^c_R \gamma^\mu \ell_L \tilde{V}_{2 \mu}
+ \mbox{ h.c.} \nonumber~,
%\\
%& + & \mbox{ h.c.}
\end{eqnarray}
where the \boldmath$\sigma$'s\unboldmath\ are Pauli matrices,
while $q_L$ and $\ell_L$ are the $SU(2)_L$ quark and lepton doublets
and $u_R$, $d_R$, $\ell_R$ are the corresponding singlets.
The subscripts of the \lk s
indicates the size of the $SU(2)_L$ representation
they belong to.
The $R$- and $S$-type \lk s are spacetime scalars,
whereas the $U$ and $V$ are vectors.
All \lk s carry a fermion number $F=3B+L=0,2$
and family and colour indices are implicit.
These quantum numbers are summarized in Table~\ref{t}.

\setlength{\arraycolsep}{4mm}
\renewcommand{\arraystretch}{2}
\newcommand{\sd}[2]{\raisebox{-1.5ex}{\shortstack[r]{$\strut#1$\\$\strut#2$}}}
\newcommand{\st}[3]{\raisebox{-3ex}{\shortstack[r]{$\strut#1$\\$\strut#2$\\$\strut#3$}}}
\newcommand{\stl}[3]{\raisebox{-3ex}{\shortstack[l]{$\strut#1$\\$\strut#2$\\$\strut#3$}}}

\begin{table}[htb]
\begin{center}
$
\begin{array}{|c||c|c|c|c|c|l|}
  \hline
  & J & F & T & T_3 & Q & \multicolumn{1}{c|}{\mbox{couples to}} \\
  \hline\hline
         S_1 & 0 & 2 &   0 &              0 &           -1/3 & \mbox{\normalsize$e_Lu_L~~e_Ru_R~~\nu_Ld_L$} \\
  \hline
  \tilde S_1 & 0 & 2 &   0 &              0 &           -4/3 & \mbox{\normalsize$e_Rd_R$} \\
  \hline
         R_2 & 0 & 0 & 1/2 & \sd{-1/2}{1/2} & \sd{-5/3}{-2/3} & \sd{\mbox{\normalsize$e_R\bar u_L~~e_L\bar u_R$}}{\mbox{\normalsize$e_R\bar d_L~~\nu_L\bar u_R$}} \\
  \hline
  \tilde R_2 & 0 & 0 & 1/2 & \sd{-1/2}{1/2} & \sd{-2/3}{1/3} & \sd{\mbox{\normalsize$e_L\bar d_R$}}{\mbox{\normalsize$\nu_L\bar d_R$}} \\
  \hline
         S_3 & 0 & 2 &   1 & \st{-1}{0}{1} & \st{-4/3}{-1/3}{2/3} & \stl{\mbox{\normalsize$e_Ld_L$}}{\mbox{\normalsize$e_Lu_L~~\nu_Ld_L$}}{\mbox{\normalsize$\nu_Lu_L$}} \\
  \hline
         U_1 & 1 & 0 &   0 &              0 &           -2/3 & \mbox{\normalsize$e_L\bar d_L~~e_R\bar d_R~~\nu_L\bar u_L$} \\
  \hline
  \tilde U_1 & 1 & 0 &   0 &              0 &           -5/3 & \mbox{\normalsize$e_R\bar u_R$} \\
  \hline
         V_2 & 1 & 2 & 1/2 & \sd{-1/2}{1/2} & \sd{-4/3}{-1/3} & \sd{\mbox{\normalsize$e_Rd_L~~e_Ld_R$}}{\mbox{\normalsize$e_Ru_L~~\nu_Ld_R$}} \\
  \hline
  \tilde V_2 & 1 & 2 & 1/2 & \sd{-1/2}{1/2} & \sd{-1/3}{2/3} & \sd{\mbox{\normalsize$e_Lu_R$}}{\mbox{\normalsize$\nu_Lu_R$}} \\
  \hline
         U_3 & 1 & 0 &   1 & \st{-1}{0}{1} & \st{-5/3}{-2/3}{1/3} & \stl{\mbox{\normalsize$e_L\bar u_L$}}{\mbox{\normalsize$e_L\bar d_L~~\nu_L\bar u_L$}}{\mbox{\normalsize$\nu_L\bar d_L$}} \\
  \hline
\end{array}
$
\end{center}
\caption{\footnotesize
  Main quantum numbers of the \lk s.
}
\label{t}
\end{table}
\renewcommand{\arraystretch}{1}

The best present bounds on leptoquarks
still originate from low energy experiments~\cite{sacha},
and the constraints on flavour diagonal couplings are weak, 
at best.
Some improvement is expected from HERA~\cite{brw},
but the real breakthrough should be obtained
in high energy experiments of the next generation~\cite{br,orwz}.
A particularly promising option
is a \lc\ operated in the $e^-\gamma$ mode~\cite{eg},
where single leptoquarks can be produced 
and very well studied.

\clearpage

\setlength{\arraycolsep}{2mm}
\renewcommand{\arraystretch}{.9}
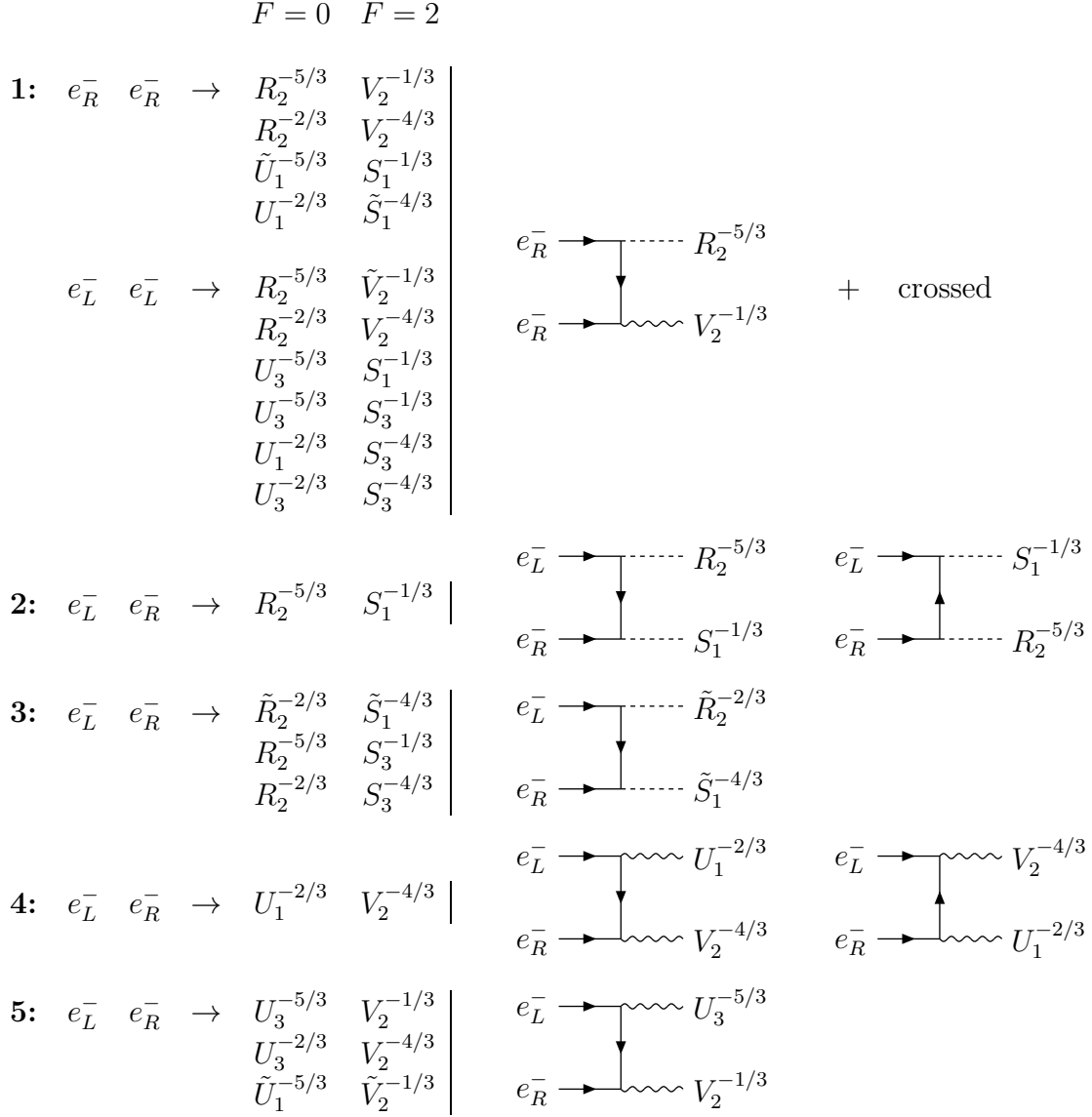
\begin{figure}[htb]
$$
\begin{array}{@{}r@{\quad}ccccc|@{\quad}l} 
&&&& F=0 & \multicolumn{2}{l}{F=2}
\\
\\\mbox{\bf1:} & e^-_R & e^-_R & \to 
   & R_2^{-5/3} & V_2^{-1/3} \\
&&&& R_2^{-2/3} & V_2^{-4/3} \\
&&&& \tilde U_1^{-5/3} & S_1^{-1/3} \\
&&&& U_1^{-2/3} & \tilde S_1^{-4/3} \\
&&&&& \\ 
& e^-_L & e^-_L & \to 
   & R_2^{-5/3} & \tilde V_2^{-1/3} 
& \begin{picture}(100,0)(-30,10)
\Text(-4,32)[rc]{$e^-_{R}$}
\Text(-4,00)[rc]{$e^-_{R}$}
\ArrowLine(00,32)(24,32)
\ArrowLine(00,00)(24,00)
\ArrowLine(24,32)(24,00)
\DashLine(24,32)(48,32){2}
\Photon(24,00)(48,00){-1}{4}
\Text(52,32)[lc]{$R_2^{-5/3}$}
\Text(52,00)[lc]{$V_2^{-1/3}$}
\end{picture}
\qquad\quad + \quad \mbox{crossed}
\\
&&&& R_2^{-2/3} & V_2^{-4/3} \\
&&&& U_3^{-5/3} & S_1^{-1/3} \\
&&&& U_3^{-5/3} & S_3^{-1/3} \\
&&&& U_1^{-2/3} & S_3^{-4/3} \\
&&&& U_3^{-2/3} & S_3^{-4/3} \\
\\
\\\mbox{\bf2:} & e^-_L & e^-_R & \to 
   & R_2^{-5/3} & S_1^{-1/3} 
& \begin{picture}(100,0)(-30,10)
\Text(-4,32)[rc]{$e^-_L$}
\Text(-4,00)[rc]{$e^-_R$}
\ArrowLine(00,32)(24,32)
\ArrowLine(00,00)(24,00)
\ArrowLine(24,32)(24,00)
\DashLine(24,32)(48,32){2}
\DashLine(24,00)(48,00){2}
\Text(52,32)[lc]{$R_2^{-5/3}$}
\Text(52,00)[lc]{$S_1^{-1/3}$}
\end{picture}
\qquad
\begin{picture}(100,0)(-30,10)
\Text(-4,32)[rc]{$e^-_L$}
\Text(-4,00)[rc]{$e^-_R$}
\ArrowLine(00,32)(24,32)
\ArrowLine(00,00)(24,00)
\ArrowLine(24,00)(24,32)
\DashLine(24,32)(48,32){2}
\DashLine(24,00)(48,00){2}
\Text(52,00)[lc]{$R_2^{-5/3}$}
\Text(52,32)[lc]{$S_1^{-1/3}$}
\end{picture}
\\
\\
\\\mbox{\bf3:} & e^-_L & e^-_R & \to 
   & \tilde R_2^{-2/3} & \tilde S_1^{-4/3} \\
&&&& R_2^{-5/3} & S_3^{-1/3} 
& \begin{picture}(100,0)(-30,10)
\Text(-4,32)[rc]{$e^-_L$}
\Text(-4,00)[rc]{$e^-_R$}
\ArrowLine(00,32)(24,32)
\ArrowLine(00,00)(24,00)
\ArrowLine(24,32)(24,00)
\DashLine(24,32)(48,32){2}
\DashLine(24,00)(48,00){2}
\Text(52,32)[lc]{$\tilde R_2^{-2/3}$}
\Text(52,00)[lc]{$\tilde S_1^{-4/3}$}
\end{picture}
\\
&&&& R_2^{-2/3} & S_3^{-4/3} \\
\\
\\\mbox{\bf4:} & e^-_L & e^-_R & \to 
   & U_1^{-2/3} & V_2^{-4/3} 
& \begin{picture}(100,0)(-30,10)
\Text(-4,32)[rc]{$e^-_L$}
\Text(-4,00)[rc]{$e^-_R$}
\ArrowLine(00,32)(24,32)
\ArrowLine(00,00)(24,00)
\ArrowLine(24,32)(24,00)
\Photon(24,32)(48,32){1}{4}
\Photon(24,00)(48,00){-1}{4}
\Text(52,32)[lc]{$U_1^{-2/3}$}
\Text(52,00)[lc]{$V_2^{-4/3}$}
\end{picture}
\qquad
\begin{picture}(100,0)(-30,10)
\Text(-4,32)[rc]{$e^-_L$}
\Text(-4,00)[rc]{$e^-_R$}
\ArrowLine(00,32)(24,32)
\ArrowLine(00,00)(24,00)
\ArrowLine(24,00)(24,32)
\Photon(24,32)(48,32){1}{4}
\Photon(24,00)(48,00){-1}{4}
\Text(52,00)[lc]{$U_1^{-2/3}$}
\Text(52,32)[lc]{$V_2^{-4/3}$}
\end{picture}
\\
\\
\\\mbox{\bf5:} & e^-_L & e^-_R & \to 
   & U_3^{-5/3} & V_2^{-1/3} \\
&&&& U_3^{-2/3} & V_2^{-4/3} 
& \begin{picture}(100,0)(-30,10)
\Text(-4,32)[rc]{$e^-_L$}
\Text(-4,00)[rc]{$e^-_R$}
\ArrowLine(00,32)(24,32)
\ArrowLine(00,00)(24,00)
\ArrowLine(24,32)(24,00)
\Photon(24,32)(48,32){1}{4}
\Photon(24,00)(48,00){-1}{4}
\Text(52,32)[lc]{$U_3^{-5/3}$}
\Text(52,00)[lc]{$V_2^{-1/3}$}
\end{picture}
\\
&&&& \tilde U_1^{-5/3} & \tilde V_2^{-1/3} 
\end{array}
$$
\caption{\footnotesize
  Leptoquark production processes 
  and their typical Feynman diagrams.
}
\label{feyn}
\end{figure}
\renewcommand{\arraystretch}{1}

Leptoquark production in $e^-e^-$ scattering~\cite{bf}
is a possibility which at first sight is not particularly attractive.
Indeed,
two \lk s have to be produced in this process,
making it inadequate for discovering these particles if they are heavy.
Electron-photon~\cite{eg}, 
electron-proton~\cite{brw} or 
proton-proton~\cite{orwz} reactions
are much better suited for this.
Moreover,
the $e^-e^-$ \xs s
involve the fourth power of the couplings to electrons and quarks,
which are totally unknown and might well be small.
In contrast,
in $e^+e^-$ annihilation~\cite{br}
or photon-gluon fusion~\cite{ed},
the \xs s mainly depend on the mass and the charge of the produced \lk s
and are always large above threshold.

%\clearpage

There is, however,
one quantum number that only one of the other experiments
is able to measure well:
the fermion number $F$.
As it turns out,
no standard reaction can see a large difference 
between $F=0$ and $F=2$ \lk s,
except electron-(anti)quark fusion~\cite{brw}
which can only take place via ($F=0$) $F=2$ \lk s,
and $e^-e^-$ scattering
which can only take place if both kinds are present.
Unfortunately,
HERA can only probe light \lk s
%is only probing a small corner of the parameter space 
%left-over by low-energy experiments.
and this situation is not going to improve
unless LEP-LHC is ever turned on.
This is why the $e^-e^-$ operating mode 
can play an important role in \lk\ searches:
if a reaction takes place,
we know for sure that we have produced 
one $F=0$ and one $F=2$ \lk.

\begin{figure}[htb]
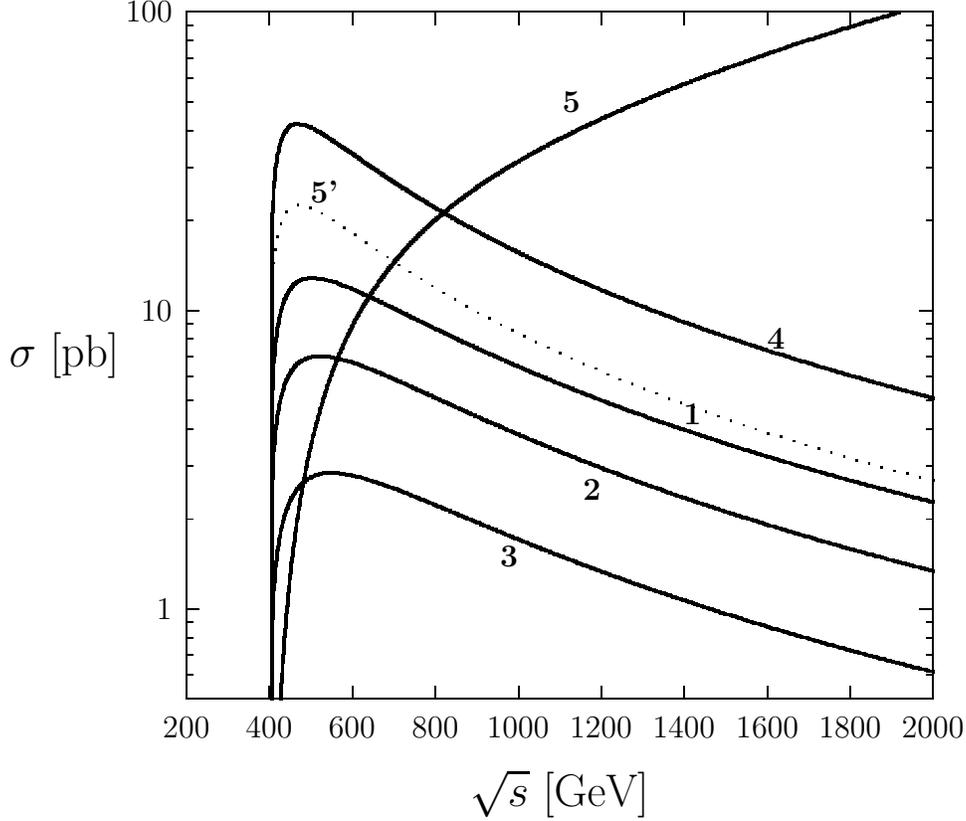

\centerline{\input en.tex}
\caption{\footnotesize
  Cross section as a function of the collider energy.
  The mass of the produced leptoquarks is 200 GeV.}
\label{eny}
\end{figure}

\begin{figure}[htb]
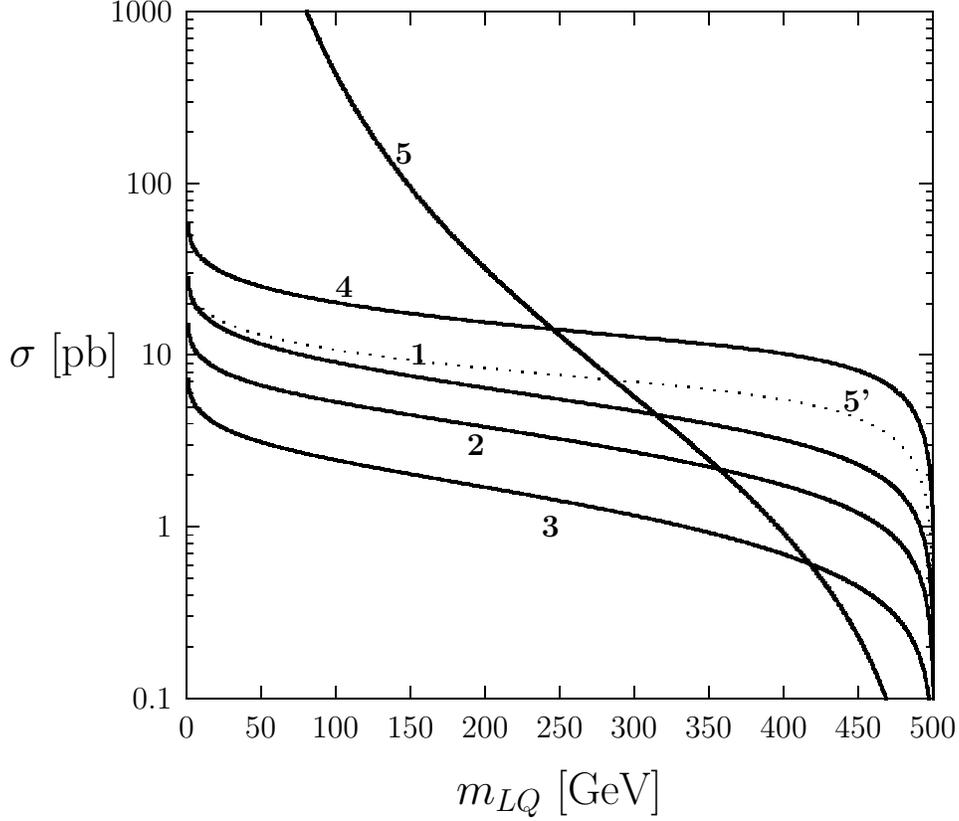

\centerline{\input ma.tex}
\caption{\footnotesize
  Cross section as a function of the mass of the produced leptoquarks.
  The collider energy is 1 TeV.}
\label{mass}
\end{figure}

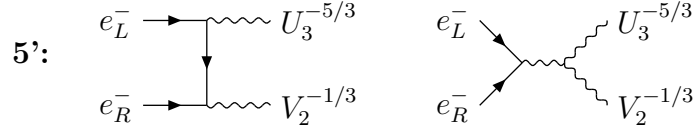
\begin{figure}[htb]
\begin{center}
{\bf5':}
\begin{picture}(100,40)(-30,16)
\Text(-4,32)[rc]{$e^-_L$}
\Text(-4,00)[rc]{$e^-_R$}
\ArrowLine(00,32)(24,32)
\ArrowLine(00,00)(24,00)
\ArrowLine(24,32)(24,00)
\Photon(24,32)(48,32){1}{4}
\Photon(24,00)(48,00){-1}{4}
\Text(52,32)[lc]{$U_3^{-5/3}$}
\Text(52,00)[lc]{$V_2^{-1/3}$}
\end{picture}
\qquad
\begin{picture}(100,40)(-30,16)
\Text(-4,32)[rc]{$e^-_L$}
\Text(-4,00)[rc]{$e^-_R$}
\ArrowLine(00,32)(16,16)
\ArrowLine(00,00)(16,16)
\Photon(16,16)(32,16){1}{3}
\Photon(32,16)(48,32){1}{4}
\Photon(32,16)(48,00){-1}{4}
\Text(52,32)[lc]{$U_3^{-5/3}$}
\Text(52,00)[lc]{$V_2^{-1/3}$}
\end{picture}
\bigskip\bigskip
\end{center}
\caption{\footnotesize
  Typical Feynman diagrams 
  involving the $s$-channel exchange of a dilepton.
}
\label{f5p}
\end{figure}

According to the lagrangian (\ref{lag}),
there are five possible types of reactions,
which we list in Fig.~\ref{feyn}.
Their total \xs s are
\bea
\sigma_1 
& = &
4G
\left[~
S + L\left( D + 2m_S^2 + 2{m_S^4 \over D} \right)
~\right] 
\\\nonumber\\
\sigma_2 
& = &
2G
\left[~
3S + L\left( D + 2{m_0^2m_2^2 \over D} \right)
~\right]
\\\nonumber\\
\sigma_3
& = &
G
\left[~
2S + LD
~\right]
\\\nonumber\\
\sigma_4 
& = &
8G
\left[~
2S + L\left( D + 2(m_0^2+m_2^2) + 2{(m_0^2+m_2^2)^2 \over D} \right)
~\right]
\\\nonumber\\
\sigma_5 
& = &
G
\left[~
{S\over6} \left( {D^2 \over m_0^2m_2^2} + 12({D\over m_0^2}+{D\over m_2^2}) 
+ 12({m_0^2\over m_2^2}+{m_2^2\over m_0^2}) - 28 \right)
 - 4LD
~\right]
\eea
where
\be
\left\{
\begin{array}{l}
\displaystyle G = {3\pi\alpha^2\over s^2} \left({\lambda\over e}\right)^4 
\\\\
\displaystyle D = s - m_0^2 - m_2^2 
\\\\
\displaystyle S = \sqrt{D^2-4m_0^2m_2^2} 
\\\\
\displaystyle L = \ln{D+S \over D-S}
\end{array}
\right.
\qquad
\left\{
\begin{array}{l}
\lambda^2 = hg = \mbox{leptoquark-lepton-quark couplings}
\\\\
m_0 = \mbox{mass of the $F=0$ leptoquark}
\\\\
m_2 = \mbox{mass of the $F=2$ leptoquark}
\\\\
m_S = \mbox{mass of the scalar leptoquark}
\end{array}
\right.
\ee
The energy and mass dependence
of these \xs s
are displayed in Figs~\ref{eny} and \ref{mass},
assuming 
\begin{itemize}
\item 100\%\ polarized beams,
\item both leptoquarks have the same common mass $m_{LQ}$ and
\item they couple to the electron and quark
  with the generic coupling $\lambda^2 = hg = e^2$.
  This is not realistic,
  but may serve as a benchmark.
\end{itemize}
The high energy and threshold behaviours of these cross sections
are the following:
\be
\begin{array}{lr|ccl}
&& \sigma_{1-4} & \propto & \displaystyle{1\over s}\ln{s\over m_0m_2} \\
&  s \gg m_0^2,m_2^2 : & \\
&& \sigma_5 & \propto & s 
\\\\
&& \sigma_{1-4} & \propto & \displaystyle\sqrt{s-(m_0+m_2)^2} \\
&  s \approx (m_0+m_2)^2 : & \\
&& \sigma_5 & \propto & \displaystyle s-(m_0+m_2)^2 
\end{array}
\ee
The pathological breaking of perturbative unitarity 
in the process {\bf5}
is due to its purely $t$-channel nature.
A similar situation is also encountered in $e^+e^-$ scattering~\cite{br}.
There are two possible cures to this:
\begin{itemize}
\item For non-gauge \lk s,
  one anyway expects other new physics effects
  to set in at some higher energy scale.
\item For gauge \lk s, though,
  one may also expect a dilepton
  to be exchanged in the $s$-channel,
  as shown in Fig.~\ref{f5p}.
  If this is the case,
  the \xs\ becomes well behaved
  as in reactions {\bf1--4}.
  It is represented by the dotted curves in Figs.~\ref{eny} and \ref{mass},
  with the dilepton mass set equal to the common \lk\ mass.
  We number this type of reaction {\bf5'}.
  Its \xs\ is
\end{itemize}
\begin{eqnarray}
\sigma_{5'} & = &
G \quad \Biggl\{ \quad
{S\over6}~ 
\biggl[ 
  62 
+ {2m_D^2 - m_0^2 \over m_2^2}
+ {2m_D^2 - m_2^2 \over m_0^2}
- {m_D^4 \over m_0^2m_2^2}
\\\nonumber
&+& {2\over s-m_D^2} \biggl(
  5m_0^2 + 5m_2^2 + 22m_D^2
- {m_D^6 \over m_0^2m_2^2}
\\\nonumber
&& \quad\qquad\qquad\qquad 
+ {9m_0^2m_D^2 - 5m_0^4 - 3m_D^4 \over m_2^2}
+ {9m_2^2m_D^2 - 5m_2^4 - 3m_D^4 \over m_2^2}
\biggr)
\\\nonumber
&-& {1\over (s-m_D^2)^2} \biggl(
  8m_0^4 + 8m_2^4 - 32m_D^4 - 32m_0^2m_D^2 - 32m_2^2m_D^2 - 18m_0^2m_2^2 
+ {m_D^8 \over m_0^2m_2^2}
\\\nonumber
&& %\quad\qquad\qquad 
+ {8m_0^4m_D^2 - 18m_0^2m_D^4 + m_0^6 + 8m_D^6 \over m_2^2}
+ {8m_2^4m_D^2 - 18m_2^2m_D^4 + m_2^6 + 8m_D^6 \over m_0^2}
\biggr)
\biggl] 
\\\nonumber
& + & 
4L \left[ D+2(m_0^2+m_2^2)+2{m_0^2m_2^2 + m_2^2m_D^2 + m_D^2m_0^2 \over s-m_D^2} \right]
\quad\Biggr\} 
\end{eqnarray}

The leptoquarks produced in the processes {\bf1}--{\bf5'}
decay with a substantial branching ratio
into a charged lepton and a jet~\cite{br}.
If the leptoquarks are family diagonal
({\em i.e.}, 
they couple to only one and the same generation of quarks and leptons),
the decay lepton is an electron.
Close to threshold,
the signature is thus
two high transverse momentum electrons-quark pairs
and no missing energy.
If we cut out the (very few) jet pairs with an invariant mass
around the $Z^0$~\cite{cor},
the only background left over originates from
the tiny quark photoproduction
$e^-e^- \to e^-e^-q\bar q$.
These events will also be removed by requiring the invariant masses
of the electron-quark pairs
to be centered around the \lk s' masses.
If the \lk s mix different families,
there can be a muon instead of an electron.
In this case, of course, 
there is no background at all.

\begin{figure}[htb]
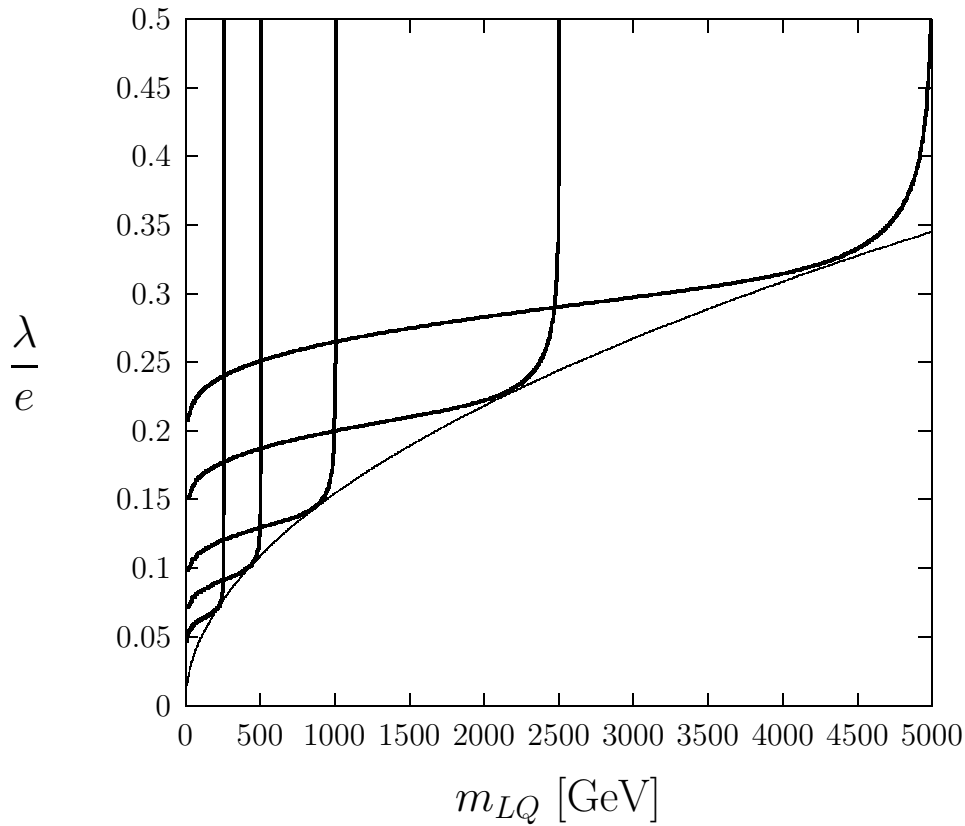

\centerline{\input lim.tex}
\caption{\footnotesize
  Loci of $\sigma_4=1$ fb,
  as a function of the leptoquark mass and coupling to fermions.
  The collider energies are .5, 1, 2, 5 and 10 TeV.
  The thinner osculating parabola is given by Eq.~(\protect\ref{lim}).}
\label{flim}
\end{figure}

To estimate the discovery potential,
we have plotted in Fig.~\ref{flim}
the boundary in the $(m_{LQ},\lambda/e)$ plane, 
below which less than 10 events 
are observed in reaction {\bf4}.
For this we consider 5 different collider energies
and assume 10 fb$^{-1}$ of accumulated luminosity.
Again the produced leptoquarks have the same mass.
In general,
the osculating curve
is approximately given by
\be
{\lambda \over e} \quad = \quad 0.35 \quad \sqrt{m_{LQ}/\mbox{TeV}} \quad 
\left(n \over A~{\cal L}/\mbox{fb}^{-1}\right)^{1/4}
\qquad \qquad 
m_{LQ} \le .43\sqrt{s}
\label{lim}~,
\ee
where 
$\lambda=\sqrt{hg}$ is the geometric mean of the leptoquarks' couplings,
$m_{LQ}$ is their common mass,
$n$ is the required number of events,
$\cal L$ is the available luminosity and
$A=6,~3,~1,~24,~12$
for reactions
{\bf1}, {\bf2}, {\bf3}, {\bf4}, {\bf5'}
respectively.

To summarize,
we have studied \lk\ production 
in the $e^-e^-$ mode of a \lc\ of the next generation.
To perform this analysis,
we have considered all types of scalar and vector \lk s,
whose interactions with leptons and quarks
conserve lepton and baryon number
and are invariant under the \sm\
$SU(3)_c \otimes SU(2)_L \otimes U(1)_Y$
gauge group.
The \sm\ backgrounds can be reduced to almost zero 
by simple kinematical cuts,
and the discovery potential is conveniently summarized
by the scaling relation Eq.~(\ref{lim}).

The $e^-e^-$ \lk\ production processes
are particularly interesting,
because to lowest order
they can only produce a pair of \lk s,
one with fermion number $F=0$ 
and the other with $F=2$.
Therefore,
the observation of such events
would demonstrate the simultaneous
existence of these two states.
Similarly,
the non-observation of this mechanism
would impose strong bounds on extensions of the \sm.
Except for electron-(anti)proton collisions,
the other standard experiments
are very weakly sensitive to this quantum number.

%\section*{Acknowledgements}
\bigskip\bigskip

Many fruitful discussions with 
Sacha Davidson, 
Paul Frampton and
Reinhold R\"uckl
are gratefully acknowledged.

%\section*{References}

\end{document}